\RequirePackage{fix-cm}

\documentclass[smallextended]{svjour3}       
\smartqed  
\usepackage{graphicx}
\usepackage{amsmath}
\usepackage{amssymb}
\usepackage{graphicx}
\usepackage{dcolumn}
\usepackage{bm}
\usepackage{setspace}
\usepackage{amsfonts}
\usepackage{rotating}
\usepackage{makeidx}
\usepackage{booktabs}
\usepackage{color}

\begin{document}

\title{Sequential state discrimination with quantum correlation \footnote{Quantum Information Processing (2018) 17:260}}


\author{Jin-Hua Zhang \and Fu-Lin Zhang\footnote{Corresponding author: flzhang@tju.edu.cn} \and Mai-Lin Liang  }


\institute{Jin-Hua Zhang \at Department of Physics, Xinzhou
Teacher's University, Xinzhou 034000, China   \at School of
Mathematical Science, Capital normal university, Beijing 100048,
China
            \and  Fu-Lin Zhang \and Mai-Lin Liang
           \at Department of Physics, School of Science, Tianjin University, Tianjin 300072, China
}

\date{Received: date / Accepted: date}

\maketitle

\begin{abstract}
The sequential unambiguous state discrimination (SSD) of two states prepared in arbitrary prior probabilities is studied, and compared with three strategies that allow classical communication.
The deviation from equal probabilities contributes to the success in all the tasks considered.
When one considers at least one of the parties succeeds, the protocol with probabilistic cloning is superior to others, which is not observed in the special case with equal prior probabilities.
We also investigate the roles of quantum correlations in SSD, and show that the procedure requires discords but rejects entanglement.
The left and right discords correspond to the part of information extracted by the first observer and the part left to his successor respectively.
Their relative difference is extended by the imbalance of prior probabilities.

\end{abstract}

\section{Introduction}
\label{intro}

The roles of quantum correlations in quantum information procedures is a fundamental problem in quantum information.
These correlations have been widely investigated in various perspectives such as quantum entanglement \cite{Horodecki09}, Bell nonlocality \cite{Bell}, and quantum discord \cite{PhysRevLett.88.017901,henderson2001classical}.
 One of the interesting findings in this field is that the
algorithm for deterministic quantum computation with one qubit
(DQC1) can surpass the performance of the corresponding classical
algorithm in the absence of entanglement between the control qubit
and a completely mixed state
\cite{lanyon2008experimental,datta2008quantum}. Thus, the
entanglement which had been regarded as the only resource for
demonstrating the superiority of quantum information processing
\cite{Horodecki09,key} is considered  to be completely unnecessary
\cite{Pang2013PRA}. The quantum discord, which gives a measurement
of the nonclassical correlations and can exist in a separable
state, is considered to be the key resource in this quantum
algorithm and has gained wide attention
\cite{modi2010unified,Bellomo2012PRA}.

Another example aided by quantum discord rather than entanglement
is the procedure of unambiguous state discrimination assisted by
an auxiliary qubit \cite{Roa2011PRL,Zhang2013SR} . Unambiguous
discrimination among linearly independent nonorthogonal quantum
states is a fundamental subject in quantum information theory
\cite{Peres1988PLA,Dies1988PLA,Bennett1992PRL,Bergou2003PRL,Pang2009PRA}.
In its simplest form, Alice prepares a qubit in one of two known
nonorthogonal states, $|\Psi_1\rangle$ and $|\Psi_2\rangle$, and
sends it to the observer Bob. Bob's task is to determine the state
he received  with no error permitted. The measurement has three
possible outcomes, $|\Psi_1\rangle$, $|\Psi_2\rangle$, and
failure, in which the last one is the price to pay for no error.
This is realized by a positive-operator-valued measurement (POVM)
on the qubit, which requires a three dimensional Hilbert space
\cite{Roa2002}.  The Hilbert space can be extended via either the
tensor product extension or the direct sum extension
\cite{Chen2007PRA,PRA2008,QINP2012} . The former is necessary when
the dimension of the measured system is fixed, e. g.  a qubit
realized by a spin-half particle.
 In such cases, Bob has to introduce an ancillary system to couple with the principal one.
 This prompts the researchers \cite{Roa2011PRL,Zhang2013SR} to study the quantum correlations (entanglement and discord) created in the discrimination process.


The work \cite{Pang2013PRA} goes even further, studying the
quantum correlations in sequential state discrimination (SSD)
presented in \cite{Bergou2013PRL}. In the protocol of SSD, another
observer Charlie will also perform an unambiguous discrimination
measurement on the same qubit after Bob's measurement. It is one
of the theories to extract information from a quantum system by
multiple observers \cite{Bergou2013PRL,Nagali2012SR,Filip2011PRA}.
The results in \cite{Pang2013PRA} demonstrate that the
entanglement is not only unnecessary for Bob's recognition, but
also an obstacle for the next observer Charlie. The left discord
of the state in Bob's hands corresponds to the information he
extracts, and the right one to the information he left to Charlie.

However, both the researches \cite{Pang2013PRA} and
\cite{Bergou2013PRL} have been limited to the special case with
equal prior probabilities. There are some critical reasons for
solving the general problem with arbitrary priors. The optimal
solution to an equal-prior problem often has a symmetric form.
We can check the robustness of  optimal solution against
variations of the priors around $1/2$ through a general
non-uniform prior result. In both the probabilistically cloning of
two pure states \cite{Yerokhin2016PRL} and sequential mixed states
discrimination \cite{Namkung2017PRA}, initial states prepared with
general non-uniform prior have demonstrated great significance.
Thus, the present study will complete the results in the general
case with arbitrary probabilities and check whether the existing
conclusions in \cite{Pang2013PRA} hold.

In the next section, we give the details of SSD with arbitrary
prior probabilities. We show the absence of entanglement is
required for SSD.
It is compared with other three protocols that allows classical communication in Sec. \ref{Other}.
The roles of quantum correlations are discussed in Sec. \ref{Discord}.
And the final section is a summary.

\section{Sequential state discrimination}\label{SSD}

\begin{figure}
\includegraphics[width=5cm]{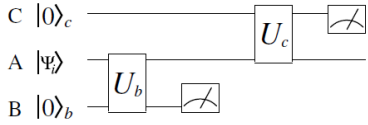} \\
 \caption{Protocol for SSD. Alice has a qubit $A$ prepared in one
 of the two nonorthogonal states $|\Psi_1\rangle$ and
 $|\Psi_2\rangle$ with prior probabilities $P_1$ , $P_2$ respectively.
After the qubit is sent to Bob, a joint unitary operation is
performed between the qubit $A$ and an auxiliary qutrit $B$,
followed by a von Neumann measurement on the qutrit. The state
discrimination is successful if the outcome is 1 (for
$|\Psi_1\rangle$) and 2 (for $|\Psi_2\rangle$), but unsuccessful
if the outcome is 0. Then the qubit in the postmeasurement state
is sent to Charlie by Bob. Charlie performs a similar joint
unitary operation $U_c$ between it and his qutrit $C$ and then
makes an optimal unambiguous discrimination measurement
\cite{Bergou2013PRL} on $C$.} \label{fig1}
\end{figure}

We now consider the procedure of SSD, which is shown in
Fig.\ref{fig1}. A qubit $A$ is prepared randomly by Alice in a
state $|\Psi_i\rangle$ with prior probabilities $P_i$, where
$i=1,2$, and $P_1+P_2=1$. Without loss of generality, we take the
overlap $s = \langle \Psi_1 | \Psi_2 \rangle$ to be a real number
($0\le s \le1$)  and $P_1\in(0,1/2]$ in the present work. Alice
sends the qubit to Bob. After performing a joint unitary
transformation \begin{math}U_b \end{math} between $A$ and his
auxiliary qutrit $B$, Bob obtains the state of the composite
system as
\begin{equation}\label{Ub1}
U_b|\Psi_i\rangle|0_b\rangle=\sqrt{q_i^b}|\chi_i\rangle|0\rangle_b+\sqrt{1-q_i^b}|\phi_i\rangle|i\rangle_b,
\end{equation}
where \{${|0\rangle_b,|1\rangle_b,|2\rangle_b}$\} is a set of
basis of the ancilla, and $|\chi_{i}\rangle$ and
$|\phi_{i}\rangle$ are pure states of $A$. Then, Bob performs a
von Neumann measurement on the qutrit with respect to the basis.
He succeeds in discrimination if the ancilla collapses to
$|1\rangle_b$ or $|2\rangle_b$, while he fails if the outcome is
$|0\rangle_b$. The average success probability of Bob can be
obtained as
\begin{equation}\label{Pb1}
P_b=P_1(1-q_1^b)+P_2(1-q_2^b).
\end{equation}

The inner product is conserved under the unitary operation. Thus,
the states $|\chi_i\rangle$ satisfy the constraint
$\sqrt{q_1^bq_2^b}\langle\chi_1|\chi_2\rangle=s$. Here, we denote
the overlap $\langle\chi_1|\chi_2\rangle=t$, with $s\leq t\leq1$.
The change from $s$ to $t$ corresponds to the information Bob
extracts from the qubit $A$ in his
measurement\cite{Pang2013PRA,Bergou2013PRL}. When $t = s$, the
overlap constraint demands $q_1^b=q_2^b=1$, which leads to the
success probability  to be zero. When $t=1$, in our following
discussion, one can find that the discrimination of the next
observer has a zero success probability. That is, all the
information encoded in qubit $A$ is extracted by Bob.

For fixed values of $s$ and $t$, the success probability $P_{b}$ of Bob can be maximized into two forms as
\begin{subequations}\label{maxb}
\begin{align}
\mathrm{(i):\ \ \ }&P_{b,{\max}}=1-2\sqrt{P_1P_2}\frac{s}{t}, \ &{\mathrm{when}}\ \frac{s^2}{s^2+t^2}\leq P_1\leq\frac{1}{2}; \\
\mathrm{(ii):\ \ \ }&P_{b,{\max}}=P_2(1-\frac{s^2}{t^2}), \
&{\mathrm{when}}\ 0< P_1< \frac{s^2}{s^2+t^2}.
\end{align}
\end{subequations}
In the above two cases, the values of the $q_1^b$ corresponding to
the optimal success probabilities are (i):
$q_1^b=\sqrt{P_2/P_1}s/t$ and (ii): $q_1^b=1$. When $q_1^b=1$, the
state $|1\rangle_b$ of $B$ is absent in Eq. (\ref{Ub1}), and
simultaneously the first term in Eq. (\ref{Pb1}) vanishes. That
is, Bob ignores $|\Psi_1\rangle$ to maximize his success
probability $P_{b}$, when $P_1$ is less than the critical value
${s^2}/{(s^2+t^2)}$.
\begin{figure}
\includegraphics[width=8cm]{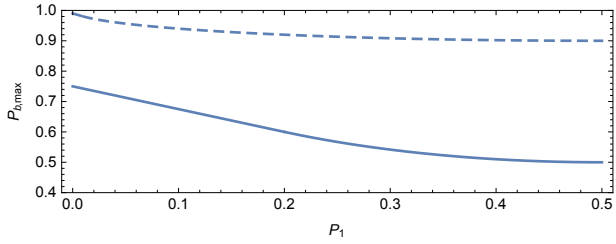} \\
 \caption{The optimal success probability  $P_{b,{\max}}$ as a function
  of the parameter $P_1$ for $s=0.05$, $t=0.06$ (solid line), $0.1$ (dotted line).} \label{fig2}
\end{figure}

As is shown in Fig. \ref{fig2}, Bob's optimal success probability
is enhanced by the deviation from equal probabilities. However, it
can't reach $1$ as $P_1$ approaches $0$. This fact is  attributed
to the requirement of no error in the task. As long as there is
even a little probability of $|\Psi_1\rangle$, the value of
$q_2^b$ is lower bounded by $s^2/t^2$ due to the overlap
constraint.

After Bob's discrimination, the qubit $A$ is sent to the second
observer Charlie, who knows Bob's protocol and performs a similar
unambiguous discrimination. A necessary condition for Charlie's
discrimination is that the states he receive are linearly
independent\cite{Pang2013PRA,Roa2002}. In Eq. (\ref{Ub1}), there
are four postmeasurement states sent to Charlie: $|\chi_1\rangle$,
$|\chi_2\rangle$, $|\phi_1\rangle$ and $|\phi_2\rangle$. In the
Hilbert space of the principal qubit, the independency requires
$|\chi_i\rangle=|\phi_i\rangle$. Thus, the transformation in Eq.
(\ref{Ub1}) should be
\begin{eqnarray}\label{Ub2}
U_b|\Psi_i\rangle|0_b\rangle=|\phi_i\rangle|\alpha_i\rangle_b,
\end{eqnarray}
where
\begin{math}|\alpha_i\rangle_b=\sqrt{q_i^b}|0\rangle_b+\sqrt{1-q_i^b}|i\rangle_b\end{math}
with \begin{math}i=1,2\end{math}. One can find that, the absence
of entanglement in the states (\ref{Ub2}) is a necessary condition
of SSD. The task of Charlie is to distinguish the states
$|\phi_1\rangle$ and $|\phi_2\rangle$ to extract the information
encoded in $|\Psi_1\rangle$ and $|\Psi_2\rangle$  by Alice.
Obviously, a necessary condition for his success is the overlap
$\langle \phi_1|\phi_2\rangle=t<1$.

Similar to Eq. (\ref{Ub2}), Charlie makes a joint unitary
operation \begin{math}U_c\end{math} between the qubit $A$ and his
auxiliary qutrit $C$, with the parameters $q_1^b$, $q_2^b$
replaced by \begin{math}q_1^c\end{math} and
\begin{math}q_2^c\end{math}. The difference is that, his two
postmeasurement states of qubit $A$ are the same, which indicates
Charlie obtains all the information left by Bob. His optimal
success probability is given by

\begin{subequations}\label{maxc}
\begin{align}
\mathrm{(i):\ \ \  }&P_{c,\max}=1-2\sqrt{P_1P_2}t,\  &{\rm{when}}\  \frac{t^2}{1+t^2}\le P_1\le\frac{1}{2};  \\
\mathrm{(ii):\ \ \ }&P_{c,\max}=P_2(1-t^2),\ &{\rm{when}}\ 0<
P_1<\frac{t^2}{1+t^2},
\end{align}
\end{subequations}
corresponding to the values (i): $q_1^c=\sqrt{P_2/P_1}t$ and (ii): $q_1^c=1$ respectively.

When $P_1\neq1/2$, although both the optimal success probabilities
$P_{b,\max}$ and $P_{c,\max}$ become piecewise functions, the
former is a monotonous increasing function of the overlap $t$ and
the later is a decreasing one. In other words, the trade-off
relation between the information extracted by Bob and Charlie
holds in the general case, which can be measured by $P_{b,\max}$
and $P_{c,\max}$ respectively.

One can obtain the success probability for both Bob and Charlie to
identify the state as
\begin{equation}\label{PSSD}
P^{SSD}=P_1(1-q_1^b)(1-q_1^c)+P_2(1-q_2^b)(1-q_2^c).
\end{equation}

Its maximum, for fixed $s$ and $P_1\leq1/2$, occurs at $t=\sqrt{s}$ and $q_1^b=q_1^c$, which indicates the equivalence between the information extracted by Bob and Charlie.
It is given by
\begin{subequations}\label{maxSSD}
\begin{align}
\mathrm{(i):\ }P^{SSD}_{\max}&=&P_1(1-q^*)^2+P_2(1-\frac{s}{q^*})^2,  \ \ \ \ \ \    {\rm{when}}  \ P_C\le P_1 \le  \frac{1}{2};  \\
\mathrm{(ii):\ } P^{SSD}_{\max}&=&P_2(1-s)^2,   \ \ \ \ \ \ \ \ \  \ \ \ \ \ \  {\rm{when}}\ 0<
P_1 \leq \min \{P_C, \frac{1}{2} \},
\end{align}
\end{subequations}
where $q^*$ satisfies  $P_1{q^* }^4- P_1{q^*}^3+P_2 s
{q^*}-P_2s^2=0$ and the critical value $P_C$   is determined by
$P_C(1-q^*)^2+(1-P_C)(1- {s}/{q^*})^2=(1-P_C)(1-s)^2$. For case
(i), the optimal success probability occurs at $q_1^b=q_1^c=q^*$,
while  $q_1^b=q_1^c=1$ for case (ii) , where Bob and Charlie
conspire to ignore the state $|\Psi_1\rangle$.

\begin{figure}
\includegraphics[width=8cm]{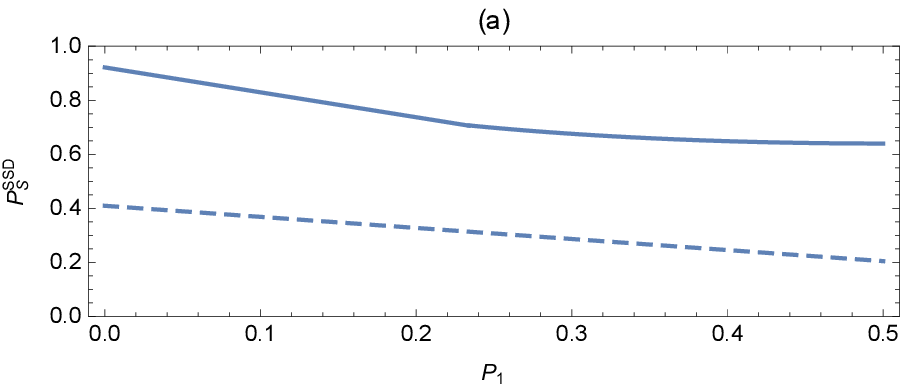} \\
\includegraphics[width=8cm]{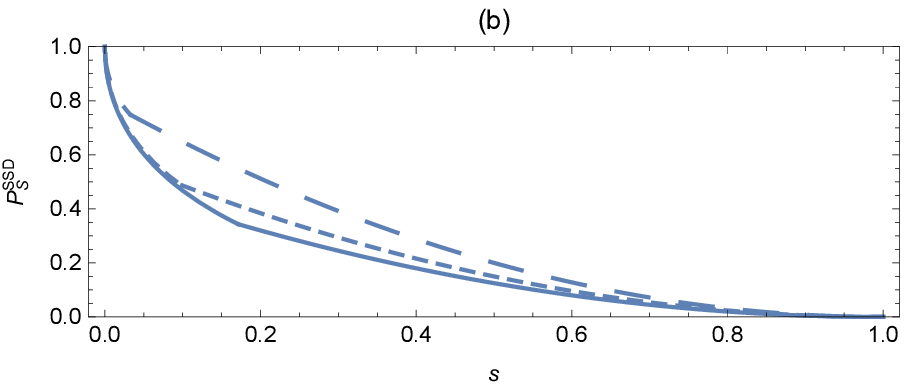} \\
 \caption{The joint optimal success probability $P^{SSD}_{\max}$ as a function
  of the parameter $P_1$ and $s$ with the other fixed which are shown in Fig (a)
  and (b) respectively.
(a): $s=0.04$ (solid line), $0.36$ (dotted line); (b): $P_1=0.5$
(solid line), $0.4$ (dotted line), $0.2$ (dashed line).}
\label{fig3}
\end{figure}

When
$s<3-2\sqrt{2}$, $P_C<1/2$, it is similar to the problem to
maximize $P_b$ or $P_c$ that the observers avoid the state with
the lower probability. However, when $s \geq 3-2\sqrt{2}$, $P_C
\geq 1/2$, the case (i) vanishes. That is, even for the
probabilities $P_1=P_2=1/2$, it is required to ignore one of the
states in the optimal solution. The phenomenon is  a symmetry
breaking due to the lack of quantum information in qubit $A$ as
pointed out in \cite{Pang2013PRA}. In Fig. \ref{fig3}, one can
find that the optimal success probability $P^{SSD}_{\max}$
decreases with the overlap $s$, and increases with the deviation
from the equal prior case. The region of case (i) is  reduced  by
$s$ and the deviation.

\section{Comparison with other protocols}\label{Other}

In this part we compare the SSD with another three strategies that
allow Bob and Charlie to communicate classically, which are
studied in \cite{Bergou2013PRL} with the equal prior
probabilities.

(1)
 Bob performs an optimal unambiguous discrimination measurement on the qubit $A$, which requires $t=1$ in Eq.(\ref{maxb}).
He sends his results to Charlie through a classical channel. If
Bob's outcome is ``failure", they end the procedure. In the
optimization, one need only consider the success of Bob, in
distinguishing the two states with overlap $s$. Consequently, the
maximal probability of both of them gaining the information sent
by Alice is given by \cite{Zhang2013SR}
\begin{subequations}\label{optimal discrimination}
\begin{align}
\mathrm{(i):\ \ \  }&P_{\max}^{(1)}=1-2\sqrt{P_1P_2}s,\ &{\rm{when}}\ \frac{s^2}{1+s^2}\leq P_1\leq\frac{1}{2};  \\
\mathrm{(ii):\ \ \ }&P_{\max}^{(1)}=P_2(1-s^2),\ &{\rm{when}}\  0<
P_1<\frac{s^2}{1+s^2}.
\end{align}
\end{subequations}
These two cases require $q_1^b=\sqrt{ {P_1}/{P_2}}s$ and $q_1^b=1$ respectively.

(2)
 Similar to the above task, Bob performs an optimal unambiguous discrimination measurement.
If the outcome is ``failure", he informs Charlie and end the
procedure. Otherwise, he sends a qubit in the state he found to
Charlie, and then Charlie performs an optimal unambiguous
discrimination on the qubit. Here, we assume that Charlie knows
the two states $|\Psi_i\rangle$ and their probabilities in Alice's
preparation.

In  Appendix \ref{Protocol2}, we show the details of optimization.
The maximal probability for both Bob and Charlie to identify the
state is a function divided in three cases as
\begin{subequations}\label{protocol2}
\begin{align}
\mathrm{(i):\ }&P_{\max}^{(2)}=(1-2\sqrt{P_1P_2}s)(1-2\sqrt{P_1^cP_2^c}s), &{\rm{when}}\ P_{c1}< P_1\leq\frac{1}{2};   \\
\mathrm{(ii):\ }&P_{\max}^{(2)}=(P_2-\sqrt{P_1P_2}s)(1-s^2),   &{\rm{when}}\ P_{c2}\leq P_1\leq P_{c1};  \\
\mathrm{(iii):\ }&P_{\max}^{(2)}=P_2(1-s^2), &{\rm{when}}\ 0 < P_1
< P_{c2},
\end{align}
\end{subequations}
where the critical probabilities are 
\begin{eqnarray}
&&P_{c1}=\frac{s^2  \left[s^4- (s^2-1 ) \sqrt{s^4-2 s^2+5}+3 \right]}{2 (s^6-s^4+3 s^2+1 )} ,\nonumber \\
&&P_{c2}=\frac{s^2}{1+s^2},
\end{eqnarray}
and $P_i^c = (P_i- \sqrt{P_1P_2} s)/(1- 2 \sqrt{P_1P_2} s)$ with
$i=1,2$. In case (i), $P_1^c$ and $P_2^c$ are the conditional
probabilities given Bob's discrimination success of
$|\Psi_1\rangle$ and $|\Psi_2\rangle$ received by Charlie. The
parameters satisfy $q_1^b=\sqrt{ {P_1}/{P_2}}s$ and $q_1^c=\sqrt{
{P_1^c}/{P_2^c}}s$. The value of $P_1^c$ decreases with the
decreasing of $P_1$ from $1/2$, and always satisfies $P_1^c \leq
P_1$. When $P_{c2}\leq P_1\leq P_{c1}$, the conditional
probability of $|\Psi_1\rangle$ is less than the value of $P_2^c$.
To maximize the total success probability, he only recognizes
state $|\Psi_2\rangle$, which has a larger prior probability. The
parameters satisfy $q_1^b=\sqrt{ {P_1}/{P_2}}s$ and $q_1^c=1$. In
case (iii), the optimal total probability requires that Bob
ignores $|\Psi_1\rangle$, \textit{i.e.} $q_1^b=1$. The conditional
probability of $|\Psi_1\rangle$ is zero. Hence, once Charlie
receives the qubit, he can learn the state in his hands being
$|\Psi_2\rangle$. This fact makes optimal probability to be
discontinuous at the point $P_1=P_{c2}$.

(3)
 Bob performs a probabilistic unitary optimal clone operation on the qubit he receives from Alice \cite{Yerokhin2016PRL,Duan1998}.
If Bob succeeds in cloning, he keeps one copy and sends the other
one to Charlie. Then, Bob and Charlie perform optimal unambiguous
discriminations to their respective qubits independently. While if
Bob's cloning fails, he will inform Charlie and end the procedure.

The maximal probability of both of their succeeding is
\begin{equation}\label{clone2}
P_{\max}^{(3)}=P_{\max}^{\rm{cl}}P_{b,\max}^{\rm{cl}}P_{c,\max}^{\rm{cl}},
\end{equation}
where $P_{\max}^{\rm{cl}}$ is the maximal success cloning
probability, and $P_{b,\max}^{\rm{cl}}=P_{c,\max}^{\rm{cl}}$ are
the optimal success probabilities of the two discriminations. The
form of $P_{b,\max}^{\rm{cl}}$ and $P_{c,\max}^{\rm{cl}}$ are the
same as Eq. (\ref{optimal discrimination}), with the probabilities
$P_i$ replaced by $P_i^{\rm{cl}}$. Here, $i=1,2$ and
$P_i^{\rm{cl}}$ are the conditional probabilities of
$|\Psi_i\rangle$ given Bob's cloning success, whose relations with
$P_i$ and $s$ are given in Appendix \ref{appa}. The form of
$P_{\max}^{\rm{cl}}$ and details to maximize the success
probability in this protocol are shown in Appendix \ref{appa}.

\begin{figure}

\includegraphics[width=8cm]{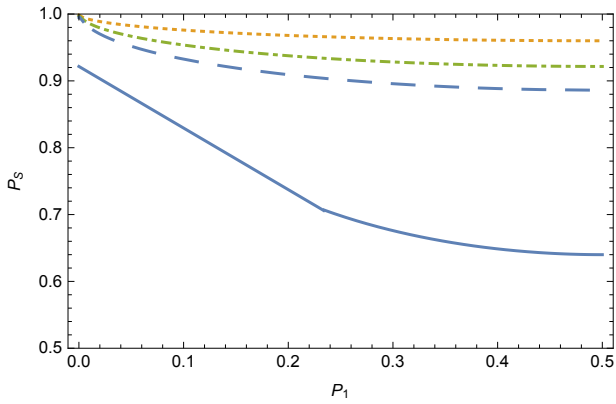} \\
 \caption{The joint optimal success probability $P_{\max}$ as a function of $P_1$ is shown for $s=0.04$ corresponding to the four
 strategies respectively. Solid line: $P_{\max}^{SSD}$;  dotted line:
 $P_{\max}^{(1)}$;
  dot-dashed line: $P_{\max}^{(2)}$; dashed line: $P_{\max}^{(3)}$.} \label{fig4}

\end{figure}


These results show that the joint optimal success probabilities of
the above four protocols have more complicated properties than the
special case with equal prior probabilities. In Fig. \ref{fig4},
one can find that as $P_1$ decreasing from $1/2$, it becomes
easier to extract the information sent by Alice in all the
strategies. And, meanwhile, their differences decreases and the
order remains unchanged.

\begin{figure}
\includegraphics[width=8cm]{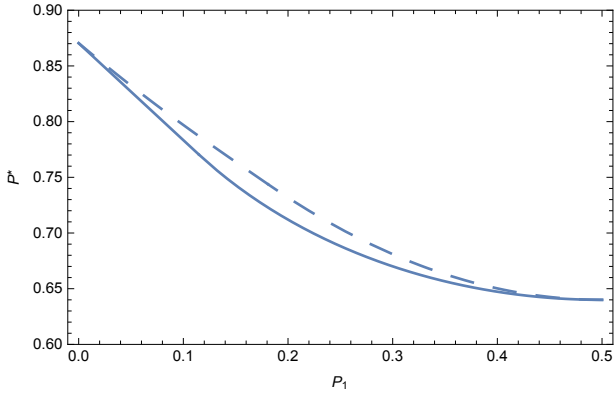} \\
 \caption{The probability $P^{*}$ for one of the two parties succeeds as a function of $P_1$
 for $s=0.36$.
 The four curves correspond to the four strategies respectively.
 Solid line: $P_{\max}^{SSD*}$, $P_{\max}^{(1)*}$, $P_{\max}^{(2)*}$; dashed line: $P_{\max}^{(3)*}$. } \label{fig5}
\end{figure}

Then we consider the probabilities $P^*$ that at least one of the
observers succeeds in identifying the states, which are shown in
Fig.\ref{fig5}. It can be noticed that, the optimal probabilities
of protocols (1), (2) and SSD  are the same, where are
$P_{\max}^{SSD*}=P_{\max}^{(1)*}=P_{\max}^{(2*)}=P_{\max}^{(1)}$.
But for the cloning protocol, the optimal probability is given by
(see Appendix \ref{appa1} for details)
\begin{subequations}\label{clone5}
\begin{align}
\mathrm{(i):\ \ \  }&P_{\max}^{(3)*}=P^{\rm{cl}}_{\max}\left(1-4P^{\rm{cl}}_1P^{\rm{cl}}_2s^2\right), \ &{\mathrm{when}}\ \frac{s^2}{1+s^2} \leq P^{\rm{cl}}_1< \frac{1}{2} ;  \\
\mathrm{(ii):\ \ \
}&P_{\max}^{(3)*}=P^{\rm{cl}}_{\max}\left[1-(P^{\rm{cl}}_1+P^{\rm{cl}}_2s^2)^2\right],
\ &{\mathrm{when}}\ 0< P^{\rm{cl}}_1< \frac{s^2}{1+s^2}.
\end{align}
\end{subequations}
As is shown in Fig. \ref{fig5}, these optimal success
probabilities $P^*$ are enhanced by the deviation from equal prior
probabilities. A difference with the existing results in
\cite{Bergou2013PRL} is that protocol (3)  is superior to the
other three strategies when $0 < P_1 < 1/2$.

Above all, the deviation from equal prior probabilities makes it
easier to gain the information sent by Alice. In the case with
general prior probabilities, the strategies that allow
communication all do better than SSD in the sense of both the
observers gaining the information. In \cite{Bergou2013PRL}, this
fact is ascribed to that SSD uses only one qubit while the others
use more. This view is supported by our results when one considers
at least one of the parties succeeds. Namely, as a new qubit is
included after Bob's cloning in protocol (3), the discriminations
of the two observers are independently from each other, while
Charlie's success depends on Bob in the other three protocols.

\section{Discords in SSD}\label{Discord}

The key step in the process of discrimination is the joint unitary
transformation between the system and  ancilla in Eq.(\ref{Ub2}),
followed by orthogonal measurements on the ancilla. This prompts
us to examine roles of quantum correlations between the principal
and the auxiliary systems in SSD with general non-uniform  prior
probabilities. Since entanglement is completely excluded by the
form of Eq. (\ref{Ub2}), we focus on quantum discords in this
part. The separable state in the discrimination of Bob can be
written as
\begin{eqnarray}\label{system state}
\rho_{AB} = P_1|\phi_1\rangle\langle\phi_1|\otimes|\alpha_1\rangle_b\langle\alpha_1| +P_2|\phi_2\rangle\langle\phi_2|\otimes|\alpha_2\rangle_b\langle\alpha_2|.
\end{eqnarray}
Discord is a kind of quantum correlation, which can exist in a
separable state. It can be considered as the part of total
correlation, measured by the quantum mutual information, which can
be disturbed by the measurements on a subsystem
\cite{henderson2001classical}. That is, there are two discords in
state $\rho_{AB}$, corresponding to the measurements on subsystem
$A$ or $B$. In the present work, we call the one with the
measurements on $A$ as left discord and denote it as $D_{BA}$,
while the other as right discord and $D_{AB}$.

The two discords of the two-rank system $\rho_{AB}$ can be derived
by using the Koashi-Winter identity \cite{Koashi2004}. Namely, one
can consider $\rho_{AB}$ as a reduced state of the tripartite
state
\begin{equation}
|\Psi\rangle=\sqrt{P_1}|\phi_1\rangle|\alpha_1\rangle_b|0\rangle_e+\sqrt{P_2}|\phi_2\rangle|\alpha_2\rangle_b|1\rangle_e,
\end{equation}
where \{$|0\rangle_e$, $|1\rangle_e$\} is the basis of a environment qubit $E$.
Then, it is directly to obtain the residual tangle $\tau_{ABE}$ of the tripartite state and the tangles between one party with the other two as \cite{Coffman2000PRA}
\begin{eqnarray}\label{tangle}
\tau_{ABE}=4P_1P_2(1-t^2)(1-r^2),&\ & \tau_{A|EB}=\tau_{A|BE}=4P_1P_2(1-t^2),\nonumber\\
\tau_{B|EA}=\tau_{B|AE}=4P_1P_2(1-r^2),&\ & \tau_{E|BA}=\tau_{E|AB}=4P_1P_2(1-t^2r^2),
\end{eqnarray}
where we set $r=s/t=\langle\alpha_1|\alpha_2\rangle=\sqrt{q_1^b q_2^b}$.
The right discord can be explicitly expressed as
\begin{equation}\label{discord}
D_{AB}=H(\tau_{B|AE})-H(\tau_{E|AB})+H(\tau_{A|BE}-\tau_{ABE}),
\end{equation}
where
\begin{eqnarray}
H(x)=-\frac{1+\sqrt{1-x}}{2}\log_2\frac{1+\sqrt{1-x}}{2}
-\frac{1-\sqrt{1-x}}{2}\log_2\frac{1-\sqrt{1-x}}{2}.
\end{eqnarray}
The left discord $D_{BA}$ can be easily obtained by interchanging
the subscripts $A$ and $B$ in Eq. (\ref{discord}). To analyze the
roles of quantum discords in SSD, we define the proportion of left
(right) discord in their total as
\begin{equation}
\tilde{D}_{\rm{left}}=\frac{D_{BA}}{D_{BA}+D_{AB}}, \ \ \ \tilde{D}_{\rm{right}}=\frac{D_{AB}}{D_{BA}+D_{AB}},
\end{equation}
and a symmetrized discord as
\begin{equation}
D_{\rm{symm}}=\sqrt{D_{BA}D_{AB}}.
\end{equation}

\begin{figure}
\includegraphics[width=8cm]{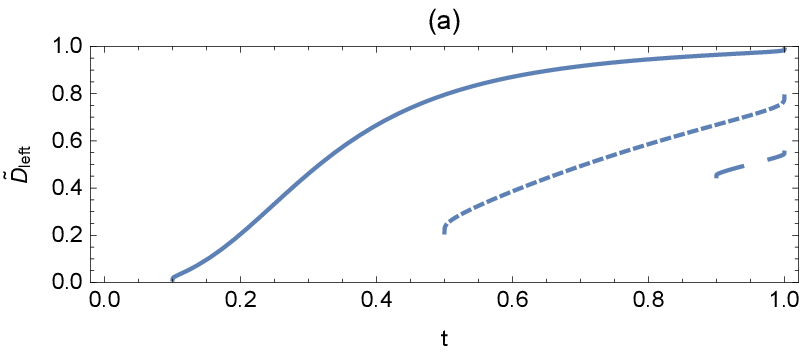} \\
\includegraphics[width=8cm]{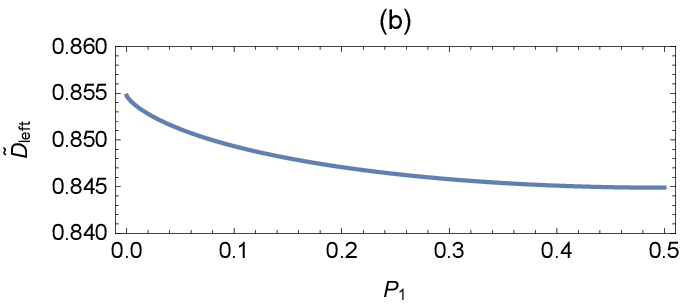} \\
\includegraphics[width=8cm]{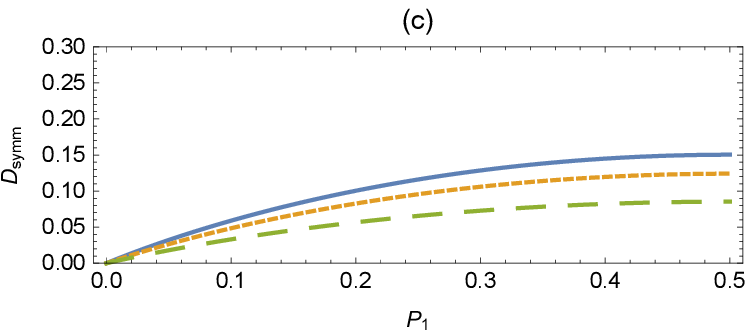} \\
 \caption{$\widetilde{D}_{left}$ as a function of the parameter
 $t$ and $P_1$ with the other variables fixed shown in (a) and (b).
 (a): $P_1=0.2$, $s=0.1$ (solid line), 0.5 (dotted line), 0.9 (dashed
 line);
(b): $s=0.1$, $t=s^{1/4}$; Fig(c) corresponds to the symmetrized
discord as a function of the prior probability $P_1$
 for $s=0.36$, $t=s^{1/2}$ (solid line), $s^{1/4}$ (dotted line) and $s^{1/8}$ (dashed line) respectively.} \label{fig6}
\end{figure}

From the form of $\rho_{AB}$ in Eq. (\ref{system state}) and the
relations in Eqs. (\ref{tangle}) and (\ref{discord}), an obvious
property can be noticed is that one can obtain $D_{BA}$ by
exchanging $r$ and $t$ in $D_{AB}$. Consequently, $D_{\rm{symm}}$
is symmetric under the exchanging of $r$ and $t$. Based on the
symmetries and the curves in Fig. \ref{fig6}, we can check the
conclusions in \cite{Pang2013PRA} and the influence of prior
probabilities on discords.

As is shown in Fig.  \ref{fig6} (a), for fixed $P_1$ and $s$, the
proportion of left discord $\tilde{D}_{\rm{left}}$ increases with
the overlap $t$. According to the mentioned symmetries,
$\tilde{D}_{\rm{right}}$ increases with $r$, and consequently
decreases with $t$. Hence, the information extracted by Bob is
positively correlated with $\tilde{D}_{\rm{left}}$, and the
information left to Charlie corresponds to
$\tilde{D}_{\rm{right}}$.

When $t=r=s^{1/2}$, the state $\rho_{AB}$ is symmetric under the
permutation of $A$ and $B$. Thus,
$\tilde{D}_{\rm{left}}=\tilde{D}_{\rm{right}}=1/2$ is independent
of the prior probabilities. When $t>s^{1/2}>r$,
$\tilde{D}_{\rm{left}}>\tilde{D}_{\rm{right}}$, and otherwise
$\tilde{D}_{\rm{left}}<\tilde{D}_{\rm{right}}$. Fig. \ref{fig6}
(b) shows a curve in the former case, where
$\tilde{D}_{\rm{left}}$ is enhanced as $P_1$ moves away from
$1/2$. According with the symmetry of $\rho_{AB}$ in Eq.
(\ref{system state}), one can learn that the deviation increases
the larger one while decreases the smaller one, and consequently
enlarges their difference.

Fig. \ref{fig6} (c) shows the symmetrized discord as a function of
$P_1$ for different values of $t$. For fixed values of $P_1$ and
$s$, the maximum symmetrized discord is reached at $t=r=s^{1/2}$,
where occurs the optimal joint probability for both Bob and
Charlie to identify the state. When $t=1$ or $s$, both the two
discords $D_{AB}$ and $D_{BA}$ are zero as the state $\rho_{AB}$
becomes a product state. Thus, the discords are needed to realize
the task of SSD. For fixed $t$, the value of $D_{\rm{symm}}$,
together with the difficulties of the two discriminations, are
reduced by the deviation from the equal prior case. When $P_1$
approaches $0$ or $1$, one of the two terms in $\rho_{AB}$ in Eq.
(\ref{system state}) vanishes, and consequently both  $D_{AB}$ and
$D_{BA}$ become zero.

\section{Summary and Outlook}\label{Summ}

The procedure of SSD in general case is investigated.
We focus on the influence of prior probabilities on the properties found in the existing researches \cite{Pang2013PRA,Bergou2013PRL}.
The deviation from equal probabilities represents more priori knowledge held by the observers before their measurements.
It enhances the success probabilities in all tasks considered in the present work.

In the cases with general prior probabilities, the optimal success
probabilities have more complicated details than the special cases
with equal prior probabilities. In the sense of both the observers
succeeding, all the three strategies that allow classical
communication do better than SSD. This is consistent with the
existing result, and is ascribed to that SSD uses only one qubit
while the others use more in \cite{Bergou2013PRL}. The imbalance
of prior probabilities leads to that the protocol (3) is superior
to others when one considers at least one of the parties succeeds.
This result is an evidence of the mentioned viewpoint in
\cite{Bergou2013PRL}, since a new qubit is introduced making the
two discriminations in protocol (3) to be independent from each
other.

The procedure for both Bob and Charlie to recognize the states
requires the absence of entanglement in Bob's system-ancilla
state. Quantum discords are necessary for their succeeding in
discriminations. Both the left and right discords become zero when
only one of the observers is allowed to gain the information,
while their proportions in total are correlated with the
information extracted by Bob and Charlie respectively. The
symmetrized discord and joint success probability reach their
maximums simultaneously when the left discord equals the right
one. These conclusions are independent of the prior probabilities.
The imbalance, corresponding to the priori knowledge of the
observers, reduces the symmetrized discord but extends the
relative differences between the left and right discords.

Our results may be generalized in several aspects. One of these is
the SSD for $n$ $(n\geq 3)$ nonorthogonal states, $|\psi_i\rangle$
(prepared with prior probabilities $P_i$, $i=0,1...n-1$), of a $n$
dimensional quantum system. Although, it is difficult to optimize
the probability of success $P_{SSD}$, the conclusion about
entanglement still hold true, which is required by the
independency of the states sent to the second observer. That is,
Bob's unitary transformation remains in the form of Eq.
(\ref{Ub2}), and his system-ancilla state is
\begin{equation}\label{n dimensional rho}
\rho_{AB}=\sum\limits_{i=0}^{n-1}P_i|\phi_i\rangle\langle\phi_i|\otimes|\alpha_i\rangle_b\langle\alpha_i|,
\end{equation}
where
$\langle\phi_i|\phi_j\rangle\langle\alpha_i|\alpha_j\rangle=\langle\psi_i|\psi_j\rangle$.
In addition, the ancilla is required to be a $2n-1$ dimensional
system \cite{Roa2002}. There are two extremes of Bob's
transformation in  Eq. (\ref{Ub2}) corresponding to zero discord
\cite{Daki2010PRL}, one of which is identity and another is a swap
followed by a local unitary operation on the auxiliary system. In
the former case, no information is  extracted by Bob, and in the
latter no information is  sent to Charlie. This result is the same
as the two-state case, shown in  Fig. \ref{fig6}.

Another two generalizations is the extension to more than two
consecutive observers and the one to mixed states. The
optimizations in both cases are solved partly very recently
\cite{Namkung2017PRA,Hillery2017JPA}, which are studied in POVM
formalism. Besides analyzing the more general success
probabilities, it is directly to describe these results via the
Neumark formalism and study the roles of correlations in the
procedures.

\section*{Acknowledgment}
This work is supported by NSF of China (Grant No.11675119, No.
11575125, No.11105097).





\appendix

\section{Calculations for protocol (2) that allows classical communication}\label{Protocol2}
 The optimization of success probability for both Bob and Charlie
 to succeed in identifying the state can be written as
\newline
\begin{equation}
{\rm{maximize:}}\
P^{(2)}=[P_1(1-q_1^b)+P_2(1-q_2^b)][P_1'(1-q_1^c)+P_2'(1-q_2^c)],
\end{equation}

\begin{eqnarray}\label{conditional}
{\rm{subject\ to:}}\
P_i'=\frac{P_i(1-q_i^b)}{P_1(1-q_1^b)+P_2(1-q_2^b)},\ i=1,2,\
q_1^cq_2^c=q_1^bq_2^b=s^2, \nonumber\\
 q_1^b,q_2^b,q_1^c,q_2^c\in[s^2,1], \ P_1\in(0,1/2].
 \end{eqnarray}

The values of $P_1'$ and $P_2'$ are derived as

\begin{subequations}\label{conditionalP}
\begin{align}
\mathrm{(i):\ \ \  }&P_1'=P_1^c,\ P_2'=P_2^c,\ &{\rm{when}}\  \frac{s^2}{1+s^2}\le P_1\le\frac{1}{2};  \\
\mathrm{(ii):\ \ \ }&P_1'=0,\ P_2'=1,\ &{\rm{when}}\
0<P_1<\frac{t^2}{1+t^2}.
\end{align}
\end{subequations}

The case (i) in Eq.(\ref{conditionalP}a) is divided into two
subcases: (ia) $\frac{s^2}{1+s^2}<P_1'\leq\frac{1}{2}$ and (ib)
$0\leq P_1'\leq\frac{s^2}{1+s^2}$ which correspond to the results
in Eq.(\ref{protocol2}a), (\ref{protocol2}b) respectively. The
corresponding critical values $P_{c1}$ in Eq.(\ref{protocol2}a)
and Eq.(\ref{protocol2}b) can be acquired after solving the
equation which satisfy the successive boundary condition.

For case (ii) in Eq.(\ref{conditionalP}b), Bob gets optimized
success probability for $q_1^b=1$. Then, for the next observer
Charlie, the conditional probability is found to be $0$ ($P_1'=0$)
according to Eq.(\ref{conditional}) and the state $|\psi_1\rangle$
is completely impossible to appear. Charlie can succeed in
identifying the state with $100\%$ probability because he has
learned that his state is actually $|\psi_2\rangle$. Thus, the
results in Eq.(\ref{protocol2}c) are obtained.

\section{Calculations for protocol (3) where probabilistic cloning occurs}\label{appa}

Bob's unitary cloning operation is given by \cite{Yerokhin2016PRL}
\begin{equation}\label{clone1}
U(|\Psi_i\rangle)|0\rangle)=\sqrt{\gamma_i}|\Psi_i\rangle|\Psi_i\rangle|\lambda_i\rangle+\sqrt{1-\gamma_i}|\beta\rangle|\beta\rangle|\lambda_0\rangle,
i=1,2,
\end{equation}
where $|0\rangle$ is a initialized state of the ancillas and
$|\lambda_i\rangle$, $|\lambda_0\rangle$ are orthogonal states of
the flag associated with successful cloning and failure cloning
respectively. $\gamma_i$ is the success probability of the cloning
for the state $|\Psi_i\rangle$ and $|\beta\rangle$ is a genetic
failure state.

Thus we can get an optimized successful cloning probability as

\begin{equation}
{\rm{maximize:}}\ P^{\rm{cl}}=P_1\gamma_1+P_2\gamma_2
\end{equation}

\begin{equation}\label{clone4}
{\rm{subject\ to:}}\quad
s=\sqrt{\gamma_1\gamma_2}s^2\langle\lambda_1|\lambda_2\rangle+\sqrt{(1-\gamma_1)(1-\gamma_2)}.
\end{equation}
according to Eq.(\ref{clone1}), where
$|\lambda_1\rangle=|\lambda_2\rangle$ is required for optimal
cloning \cite{Yerokhin2016PRL}.

If we set $\sin{\theta_i}=\sqrt{1-\gamma_i}$ ($i=1,2$) for
$0\leq\theta_i\leq\pi/2$, the variables
$x=\cos(\theta_1+\theta_2)$, $y=\cos(\theta_1-\theta_2)$ are
further introduced. Eq.(\ref{clone4}) is equivalent to
$2s=(1+s^2)y-(1-s^2)x$. And then we find an intermediate parameter
$\omega$ which satisfies
\begin{equation}
x=\frac{1-(1+s^2)\omega}{s}, y=\frac{1-(1-s^2)\omega}{s}.
\end{equation}
The range of the parameter $\omega$ is given in Eq.(\ref{range}).
It's found that
\begin{equation}
\gamma_i=\frac{1}{2}[1+xy+(-1)^i\sqrt{(1-x^2)(1-y^2)}].
\end{equation}
To seek the optimal value $P_{\max}^{\rm{cl}}$, the following
equation should be satisfied
$(P_{\max}^{\rm{cl}})'=\frac{dP_{\max}^{\rm{cl}}}{d\omega}=0$.
This equation is equivalent to $P_1\gamma'_1+(1-P_1)\gamma'_2=0$,
thus the following results are obtained

\begin{equation}
P_1=\frac{\gamma'_2}{\gamma'_2-\gamma'_1},
P_{\max}^{\rm{cl}}=\frac{\gamma'_2\gamma_1-\gamma'_1\gamma_2}{\gamma'_2-\gamma'_1}.
\end{equation}
where
\begin{equation}
\gamma'_i=\frac{d\gamma_i}{d\omega}=\frac{\sqrt{\gamma_i(1-\gamma_i)}}{s}[-\frac{1+s^2}{\sqrt{1-x^2}}+(-1)^i\frac{1-s^2}{\sqrt{1-y^2}}].
\end{equation}

And then, the conditional probabilities $P_i^{\rm{cl}}$ ($i=1,2$)
of $|\Psi_i\rangle$ for the following two discriminations can be
obtained as
$P_i^{\rm{cl}}=\frac{P_i\gamma_i}{P_1\gamma_1+P_2\gamma_2}$.
Hence, for the optimized successful cloning probability, $P_i$,
$P_i^{\rm{cl}}$, $P_{\max}^{\rm{cl}}$, $P_{b,\max}^{\rm{cl}}$ and
$P_{c,\max}^{\rm{cl}}$ are all obtained as parametric functions of
$\omega$ with the range
\begin{equation}\label{range}
\omega_1\leq\omega\leq\omega_2, \omega_1=\frac{1}{1+s},
\omega_2=\frac{1}{1+s^2},
\end{equation}
where $\omega_1$ and $\omega_2$ correspond to the cases for
$P_1=P_2=\frac{1}{2}$ and $P_1=0$ respectively.

At last, the optimal success probability for both Bob and Charlie
to identify the state is obtained as

\begin{equation}
{\rm{maximize:}}\
P^{(3)}=P_{\max}^{\rm{cl}}[P_1^{\rm{cl}}(1-q_1^b)+P_2^{\rm{cl}}(1-q_2^b)][P_1^{\rm{cl}}(1-q_1^c)+P_2^{\rm{cl}}(1-q_2^c)]
\end{equation}

\begin{equation}
{\rm{subject\ to:}}\ q_1^cq_2^c=q_1^bq_2^b=s^2, \
q_1^b,q_2^b,q_1^c, q_2^c\in[s^2,1]
\end{equation}.

Thus, we can acquire the results in Eq.(\ref{clone2})
analytically.

\section{Optimal probability for at least one of Bob and Charlie succeeding in identifying the states}\label{appa1}

It is obvious that the optimized probability $P_{\max}^*$ for one
of their succeeding in discrimination for protocol (1) and (2) is
equivalent to the results in Eq.(\ref{optimal discrimination}).
For SSD protocol, we can obtain the optimization as

\begin{equation}
{\rm{maximize:}}\ P^{SSD*}=P_1(1-q_1^bq_1^c)+P_2(1-q_2^bq_2^c)
\end{equation}.

\begin{eqnarray}
{\rm{subject\ to:}}\ P_1\in(0,1/2],\ q_1^bq_2^b=s^2/t^2,\
q_1^cq_2^c=t^2,
 \nonumber\\ q_1^b,q_2^b\in[s^2/t^2,1],\ q_1^c,q_2^c\in[t^2,1].
\end{eqnarray}
Thus, this result is also equal to $P_{\max}^{(1)}$. For protocol
(3), the maximal probability is derived as

\begin{equation}
{\rm{maximize:}}\
P^{(3)*}=1-(P_1^{\rm{cl}}q_1^b+P_2^{\rm{cl}}q_2^b)(P_1^{\rm{cl}}q_1^c+P_2^{\rm{cl}}q_2^c)
\end{equation}

\begin{equation}
{\rm{subject\ to:}}\ q_1^bq_2^b=q_1^cq_2^c=s^2,\ q_1^b, q_2^b,
q_1^c, q_2^c\in[s^2,1],\ P_1\in(0,1/2]
\end{equation}
Thus, the result in Eq.(\ref{clone5}) can be easily obtained.
\end{document}